\documentclass{PoS}
\usepackage{amssymb,amsfonts,amsmath}
\usepackage{amsfonts}
\usepackage{amsmath}
\usepackage{graphicx}
\usepackage{dcolumn}
\usepackage{bm}
\usepackage{textcomp}
\usepackage{amssymb}%
\title{Masses of light and heavy mesons in a $U(4)_r \times U(4)_l$ linear sigma model}

\ShortTitle{Masses of light and heavy mesons in a $U(4)_r \times
U(4)_l$ linear sigma model}

\author{\speaker{Walaa I.\ Eshraim}\thanks{In collaboration with Francesco Giacosa and Dirk H.\ Rischke}\\
        {Institute for Theoretical Physics, Goethe University, Max-von-Laue-Str. 1,
\\60438 Frankfurt am Main, Germany}\\
        E-mail: \email{weshraim@th.physik.uni-frankfurt.de}}

\abstract{We extend the three-flavor linear sigma model with
(axial-)vector mesons to four flavors. We compute the masses of (pseudo)scalar and
(axial-)vector mesons including open and hidden charmed mesons as
well as weak decay constants. The results are in good
agreement with experimental data.}

\FullConference{QCD-TNT-III-From quarks and gluons to hadronic matter: A bridge too far?,\\
        2-6 September, 2013\\
        European Centre for Theoretical Studies in Nuclear Physics and Related Areas (ECT*), Villazzano, Trento (Italy)}

\begin{document}

\section{Introduction}

For $N_f$ massless quark flavors, quantum chromodynamics
(QCD), the fundamental theory of the strong interaction, has a
global chiral $U(N_f)_r\times U(N_f)_l= SU(N_f)_r \times SU(N_f)_l
\times U(1)_V \times U(1)_A $ symmetry, where $V=r+l$, and
$A=l-r$. The $U(1)_V$ symmetry corresponds to baryon number
conservation. Effective models in the linear \cite{Schwinger1} and
nonlinear \cite{Schwinger2} realization of the chiral symmetry are
widely used to investigate the low-energy sector of the
strong interaction, e.g.\ vacuum
properties of hadrons \cite{Gasiorowicz, Meissner}. The nonlinear
realization (the so-called nonlinear sigma model) contains only
the lightest degrees of freedom, the
pseudoscalar mesons. It forms the basis of chiral perturbation
theory as shown in Ref.\ \cite{Gasser}. The linear representation
of chiral symmetry (the so-called linear sigma model) contains
both scalar and pseudoscalar degrees of freedom. (Axial-)vector
mesons can also be included in the model \cite{denis, Gallas, Ko, Urban}.

The extended Linear Sigma Model (eLSM) has been
successfully used to study the vacuum phenomenology of the nonets
of (pseudo)scalar, (axial-)vector, and tensor mesons, as shown in
Refs.\ \cite{Gasser, denis, Gallas, Ko, Urban, Lenaghan, Kovacs,
Bando, Parganlija1, stani, denisthesis} for non-strange hadrons
($N_f=2$), and Refs.\ \cite{denisthesis, three flavor, dick} for
strange hadrons ($N_f=3$).
The eLSM emulates the global symmetries of the QCD Lagrangian; the
global chiral symmetry (which is exact in the chiral limit),
the discrete C, P, and T symmetries, and the classical dilatation
(scale) symmetry. When working with colorless hadronic degrees of
freedom, the local color symmetry of QCD is automatically
preserved. In QCD (and thus also in the eLSM) the global chiral symmetry is explicitly
broken by non-vanishing quark masses and quantum effects
\cite{Hooft}, and spontaneously by a non-vanishing expectation
value of the quark condensate. The dilatation symmetry is broken
explicitly by the logaritmic term of the dilaton potential, by the mass terms, and by the $U(1)_A$ anomaly.

In the present work, we outline the extension of the eLSM
from the three-flavor case ($N_f=3$) to the four-flavor
case ($N_f=4$) which includes charm degrees of freedom. In
nature, this symmetry is strongly explicitly broken by the large
charm quark mass. Nevertheless, it is still of principle interest to see how
a linear sigma model fares in describing charmed hadron vacuum properties.
As we shall see, this works surprisingly well.

Although the present work represents a straightforward
implementation of the principles of the linear realization of
chiral symmetry, this is the first time that all these degrees of
freedom are considered within a single linear chiral framework
which includes twelve new charmed mesons in addition to the
nonstrange-strange sector. The new charmed mesons of lowest mass,
the $D,\, D_S  $, and the higher mass $\eta_C$, are
quark-antiquark spin-singlet states with quantum number
$J^{pc}=0^{-+}$, i.e., pseudoscalar mesons. The scalar mesons
$D^{*}_0,\, D^{*}_{S0}$, and $\chi_{C0}$ are spin-singlet states
with $J^p=0^{++}$. The vector mesons $D^{*}, \, D^{*}_S$, and
$J/\psi$ are quark-antiquark spin triplets with $J^{pc}=1^{--}$ .
The axial-vector mesons $D_{1}, \, D_{S1}$, and $\chi_{C1}$ are
quark-antiquark spin triplets with $J^{pc}=1^{++}$. Most
parameters of our linear sigma model are taken directly from Ref.\ \cite{dick}
where the nonstrange-strange mesons were considered. There are three new
parameters pertaining to the charm degree of
freedom. In these proceedings, we calculate all meson masses
in the model including open and hidden charmed mesons, and the decay
constants of the pseudoscalar
$D$ and $D_S$ mesons.

These proceedings are organized as follows: in Sec.\ 2 we
present the $U(4)_R \times U(4)_L$ linear sigma model with
(axial-)vector mesons and its implications. In Sec.\ 3 we fix the parameters
and present the results, and in Sec.\ 4 we provide our conclusions and an outlook.
Our units are $\hbar=c=1$, the metric tensor is $g^{\mu\nu}={\rm diag}(+,-,-,-)$.

\section{The $U(4)_r \times U(4)_l$ linear sigma model and its implications}

In this section we extend the eLSM to the case $N_f=4$ which includes open and
hidden charmed mesons. We consider isospin multiplets as a single
degree of freedom, which gives 28 resonances for $N_f=4$: \\
(i) scalar mesons: $\sigma_N, \,\sigma_S,\, a_0,\, K^*_0,\, D^*_0,
\,D^*_{S0},\, \chi_{C0}$.\\
(ii) pseudoscalar mesons: $\eta_N,
\,\eta_S, \,\pi, \,K, \,D,\, D_S, \,\eta_C$.\\
(iii) vector mesons: $\omega_N,\, \omega_S, \,\rho, \,K^{*},\,
D^{*},\, D^{*}_{S}, \,J/\psi$.\\
(iv) axial-vector mesons: $f_{1N}, \,f_{1S},\, a_1,\, K_1,\, D_1, \,D_{S1},\, \chi_{C1}$. \\

When we assign a state from our model to a physical resonance we
assume that the resonance is a $q\overline{q}$ state. There are 16
light (i.e., with mass $\lesssim 2$ GeV) resonances as
discussed in Refs.\ \cite{denisthesis, dick}, and twelve new heavy resonances.
We include all of these mesons in our model by introducing
$4\times4$ matrices as follows:\\
(i) The multiplet of the scalar, $S_i$, and the pseudoscalar,
$P_i$, quark-antiquark states:
\begin{equation}\label{3}
\Phi=\sum_{i=0}^{15}(S_i+i\,P_i)\,T_i=\frac{1}{\sqrt{2}}
\left(%
\begin{array}{cccc}
  \frac{(\sigma_{N}+a^0_{0})+i(\eta_N +\pi^0)}{\sqrt{2}} & a^{+}_{0}+i \pi^{+} & K^{*+}_{0}+iK^{+} & D^{*0}_0+iD^0 \\
  a^{-}_{0}+i \pi^{-} & \frac{(\sigma_{N}-a^0_{0})+i(\eta_N -\pi^0)}{\sqrt{2}} & K^{*0}_{0}+iK^{0} & D^{*-}_0+iD^{-} \\
  K^{*-}_{0}+iK^{-} & \overline{K}^{*0}_{0}+i\overline{K}^{0} & \sigma_{S}+i\eta_{S} & D^{*-}_{S0}+iD^{-}_S\\
  \overline{D}^{*0}_0+i\overline{D}^0 & D^{*+}_0+iD^{+} & D^{*+}_{S0}+iD^{+}_S & \chi_{C0}+i\eta_C\\
\end{array}%
\right),
\end{equation}
where $T_i\;(i=0,...,15)$ denote the generators of $U(4)$.\\
(ii) The left-handed and right-handed matrices containing the
vector, $V_i^\mu$, and axial-vector, $A_i^\mu$, degrees of
freedom:
\begin{equation}\label{4}
L^\mu=\sum_{i=0}^{15}(V_i^\mu+i\,A^\mu_i)\,T_i=\frac{1}{\sqrt{2}}
\left(%
\begin{array}{cccc}
  \frac{\omega_N+\rho^{0}}{\sqrt{2}}+ \frac{f_{1N}+a_1^{0}}{\sqrt{2}} & \rho^{+}+a^{+}_1 & K^{*+}+K^{+}_1 & D^{*0}+D^{0}_1 \\
  \rho^{-}+ a^{-}_1 &  \frac{\omega_N-\rho^{0}}{\sqrt{2}}+ \frac{f_{1N}-a_1^{0}}{\sqrt{2}} & K^{*0}+K^{0}_1 & D^{*-}+D^{-}_1 \\
  K^{*-}+K^{-}_1 & \overline{K}^{*0}+\overline{K}^{0}_1 & \omega_{S}+f_{1S} & D^{*-}_{S}+D^{-}_{S1}\\
  \overline{D}^{*0}+\overline{D}^{0}_1 & D^{*+}+D^{+}_1 & D^{*+}_{S}+D^{+}_{S1} & J/\psi+\chi_{C1}\\
\end{array}%
\right)^\mu,
\end{equation}
$$\\$$
\begin{equation}\label{5}
R^\mu=\sum_{i=0}^{15}(V_i^\mu-i\,A^\mu_i)\,T_i=\frac{1}{\sqrt{2}}
\left(%
\begin{array}{cccc}
  \frac{\omega_N+\rho^{0}}{\sqrt{2}}- \frac{f_{1N}+a_1^{0}}{\sqrt{2}} & \rho^{+}-a^{+}_1 & K^{*+}-K^{+}_1 & D^{*0}-D^{0}_1 \\
  \rho^{-}- a^{-}_1 &  \frac{\omega_N-\rho^{0}}{\sqrt{2}}-\frac{f_{1N}-a_1^{0}}{\sqrt{2}} & K^{*0}-K^{0}_1 & D^{*-}-D^{-}_1 \\
  K^{*-}-K^{-}_1 & \overline{K}^{*0}-\overline{K}^{0}_1 & \omega_{S}-f_{1S} & D^{*-}_{S}-D^{-}_{S1}\\
  \overline{D}^{*0}-\overline{D}^{0}_1 & D^{*+}-D^{+}_1 & D^{*+}_{S}-D^{+}_{S1} & J/\psi-\chi_{C1}\\
\end{array}%
\right)^\mu.
\end{equation}
The assignment of fields in the $N_f=3$ sector to physical
resonances is the following \cite{dick, GiacosaL, Amsler, Eshraim}:
the fields $\overrightarrow{\pi}$ and $K$ represent the pions and
kaons, respectively. The fields $ \omega_N,\, \omega_S, \,\,
\overrightarrow{\rho},
\, f_{1N}, \,f_{1S},$ $\overrightarrow{a}_1, \,K^*, \, K^*_0,$ and
$K_1$ are assigned to the $\omega(782),\,\phi(1020), \, \rho(770),
\, f_1(1285),\, f_1(1420), \, a_1(1260),$ $ K^*(892), \,
K^*_0(1430),$ and $K_1(1270)$, or $K_1(1400)$ mesons,
respectively. The field $\overrightarrow{a}_0$ is the physical
isotriplet state $a_0(1450)$ (the details of this assignment are
given in Ref.\ \cite{dick}).
 The bare non-strange field $\eta_N\equiv|\overline{u}u+\overline{d}d\rangle/\sqrt{2}$
and strange field $\eta_S\equiv|\overline{s}s\rangle$ mix to yield
the physical $\eta$ and $\eta'$ fields, with the pseudoscalar
mixing angle $\varphi\simeq-44.6^{\circ}$ \cite{dick, Eshraim}.
The non-strange and strange isoscalar
$\sigma_N$ and $\sigma_S$ fields mix to give the physical
isoscalar resonances $f_0(1370)$ and $f_0(1710)$, respectively \cite{denis, Eshraim}.

In the present work, we assign the additional charmed fields $
D^{*0}, \,D^{*}, \,D^{*0}_0,\,D^*_0, \, D^*_{S0}, \, \chi_{c1},\, \chi_{c0}$, and $J/\psi$ to the physical resonances
$D^*(2007)0,\, D^*(2010)^\pm, \, D^*_0(2400)^0,\, D^*_0(2400)^\pm,\, D^*_{S0}(2317),$  $\chi_{c1}(1P),\, \chi_{c0}(1P)$, and $J/\psi(1S)$, respectively. The isospin doublet $D^{0}_1$
is $D_1(2420)$. The isospin singlet $D_{S1}$ can be assigned
to two different physical resonances, $D_{S1}(2460)$ and
$D_{S1}(2536)$. Reference \cite{Molecule} found $D_{S1}(2460)$ to
be a molecule, so we assign $D_{S1}$ to $D_{S1}(2536)$.

The eLSM contains also the scalar glueball, $G$, and the
pseudoscalar glueball, $\widetilde{G}$. The scalar glueball is
included in the dilaton Lagrangian \cite{dick,nf3Giacosa,Rosenzweig}
\begin{equation}
\mathcal{L}_{dil}=\frac{1}{2}(\partial_\mu
G)^2-\frac{1}{4}\frac{m_G^2}{\Lambda^2}\left(G^4\,\rm{ln}\frac{G^2}{\Lambda^2}-\frac{G^4}{4}\right),
\end{equation}
where $\Lambda$ is a constant and is the minimum of the dilaton potential. The dilaton potential breaks the dilatation symmetry
explicitly.  The mass of the glueball is about $1.6$ GeV which obtained from lattice-QCD calculations \cite{Morningstar}.

The Lagrangian of the $N_f=4$ model with global chiral invariance
has an analogous form as the corresponding eLSM
Lagrangian for $N_f=3$ \cite{ denisthesis, three flavor,
 dick}. For a better fit to the masses, we add a new mass term $-2\,
\rm{Tr}[\varepsilon\Phi^\dagger\Phi]$. The Lagrangian then reads:
\begin{multline}\label{2.5}
\mathcal{L}=\mathcal{L}_{dil}+\rm{Tr}[(D^\mu \Phi)^\dagger (D^\mu
\Phi)]- m_0^2\left(\frac{G}{G_0}\right)^2 \rm{Tr}(\Phi^\dagger
\Phi)-\lambda_1 [\rm{Tr}(\Phi^\dagger \Phi)]^2-\lambda_2
\rm{Tr}(\Phi^\dagger
\Phi)^2\\+\rm{Tr}[H(\Phi+\Phi^\dagger)]-2\,
\rm{Tr}[\varepsilon\Phi^\dagger\Phi]+c(\rm{det}\Phi-\rm{det}\Phi^\dagger)^2+i
c_{\tilde{G}\Phi} \tilde{G}(det \Phi-det\Phi^\dagger) \\-\frac{1}{4}\rm{Tr}[(L^{\mu\nu})^2+(R^{\mu\nu})^2]+\rm{Tr}\left\{\left[\left(\frac{G}{G_0}\right)^2
\frac{m_1^2}{2}+\Delta\right]\left[(L^\mu)^2+(R^\mu)^2\right]\right\}\\-2 i g_{2}
\{\rm{Tr}(L_{\mu\nu} [L^\mu,L^\nu])+\rm{Tr}(R_{\mu\nu}
[R^\mu,R^\nu])\}+\frac{h_1}{2}\rm{Tr}(\Phi^\dagger \Phi)
\rm{Tr}[(L^\mu)^2+(R^\mu)^2]\\+h_2 \rm{Tr}[(\Phi R^\mu)^2+(L^\mu \Phi)^2]+2
h_3 \rm{Tr}(\Phi R_\mu \Phi^\dagger L^\mu),\,\,\,\,\,\,\,\,\,\,\,\,\,\,\,\,\,\,\,\,\,\,\,\,\,\,\,\,\,\,\,\,\,\,\,\,\,\,\,\,\,\,\,\,\,\,\,\,\,\,\
\end{multline}
where $\mathcal{L}_{dil}$ is the dilaton term (2.5), $D^\mu\Phi\equiv\partial^\mu\Phi-ig_1 (L^\mu \Phi-\Phi
R^\mu)$ is the covariant derivative;
$L^{\mu\nu}\equiv \partial^\mu L^\nu -
\partial^\nu L^\mu$, and $R^{\mu\nu}\equiv\partial^\mu R^\nu -
\partial^\nu R^\mu$ are the left-handed and
right-handed field strength tensors.
$H,\, \Delta$, and $\varepsilon$ are constant external fields defined as
\begin{equation}\label{2.6}
H=H_0\,T_0+H_{8}\,T_{8}+H_{15}\,T_{15}=\frac{1}{2}
\left(%
\begin{array}{cccc}
  h_{0N} & 0 & 0 &0 \\
  0 & h_{0N}& 0 & 0 \\
  0 & 0 & \sqrt{2}h_{0S} & 0 \\
  0 & 0 & 0 & \sqrt{2}h_{0C} \\
\end{array}%
\right),
\end{equation}
where $h_{0N}=const., h_{0S}=const.$, and $h_{0C}=const.,$
\\
\begin{equation}\label{2.7}
\Delta=\Delta_0\,T_0+\Delta_{8}\,T_{8}+\Delta_{15}\,T_{15}=
\left(%
\begin{array}{cccc}
  \delta_{0N} & 0 & 0 &0 \\
  0 & \delta_{0N}& 0 & 0 \\
  0 & 0 & \delta_{0S} & 0 \\
  0 & 0 & 0 & \delta_{0C} \\
\end{array}%
\right),
\end{equation}
and
\begin{equation}\label{2.8}
\varepsilon=\left(%
\begin{array}{cccc}
 0 & 0 & 0 & 0 \\
 0 & 0 & 0 & 0 \\
 0 & 0 & 0 & 0\\
 0 & 0 & 0 & \varepsilon_C\\
\end{array}%
\right),
\end{equation}
where $\delta_N\sim m_N^2, \delta_S\sim m_S^2, \delta_C\sim
m_C^2$, and $\varepsilon_C \sim m_C^2$. In our framework the
isospin symmetry is exact for up and down quarks, so in Eqs. (2.6
- 2.7) the first two diagonal elements are identical. Then, only
the scalar-isoscalar fields $\sigma_N,\, \sigma_S,\,
G$, and $\chi_{C0}$ have the quantum numbers of the vacuum and
can have nonzero expectation values. Moreover with no loss of generality, one can set $\delta_{0N}=0$.

In order to implement spontaneous symmetry breaking, we shift
$\sigma_N$ and $\sigma_S$ by their respective vacuum
expectation values $\phi_N$, $\phi_S$, and $\phi_C$ as
\begin{equation}\label{2.9}
\sigma_N\rightarrow\sigma_N+\phi_N\,\,\, \rm{and}\,\,\,
\sigma_S\rightarrow\sigma_S+\phi_S\;,
\end{equation}
as obtained in Refs.\ \cite{denis, Gallas, denisthesis}, and similarly for
$\chi_{C0}$,
\begin{equation}\label{10}
\chi_{C0}\rightarrow\chi_{C0}+\phi_C\;.
\end{equation}
The spontaneous symmetry breaking generates in $\eta_N-f_{1N}$,
$\overrightarrow{\pi}-\overrightarrow{a}_1$ [7], $\eta_S-f_{1S}$,
$K_S-K^*$, and $K-K_1$ mixing terms [11]:
\begin{multline}
-g_1\phi_N(f_{1N}^\mu\partial_\mu\eta_N+\overrightarrow{a}_1^\mu\cdot\partial_\mu\overrightarrow{\pi})
-\sqrt{2}\,g_1\phi_Sf^\mu_{1S}\partial_\mu\eta_S+ig_1(\sqrt{2}\phi_S-\phi_N)(\overline{K}\,^{*\mu0}\,\partial_\mu
K^0_S\\+K^{*\mu-}\,\partial_\mu
K^+_S)/2+ig_1(\phi_N-\sqrt{2}\phi_s)(K^{*\mu0}\,\partial_\mu\overline{K}^0_S+K^{*\mu+}\,
\partial_\mu K^{-}_S)/2-g_1(\phi_N+\sqrt{2}\,\phi_S)\\(K_1^{\mu 0}\,\partial_\mu \overline{K}^0+K_1^{\mu+}\,\partial_\mu
K^-)/2-g_1(\phi_N+\sqrt{2}\,\phi_S)(\overline{K}_1^{\mu
0}\,\partial_\mu K^0+K_1^{\mu-}\,\partial_\mu K^+)/2,
\end{multline}
respectively, as well as in $\eta_C-\chi_{C1}$, $D_S-D_{S1}$,
$D^*_{S0}-D^*_{S1}$, $D^*_0-D^*$, and $D-D_1$ mixing terms:
\begin{multline}\label{6}
-g_1\phi_C\, \chi_{C1}^\mu \, \partial_\mu \eta_C -\sqrt{2}\,g_1
\phi_S(D^{\mu-}_{S1}\,\partial_\mu
D^+_S+D^{\mu+}_{S1}\,\partial_\mu
D^-_S)/2+\sqrt{2}\,ig_1\phi_S(D^{*\mu-}_{S1}\,\partial_\mu
D^{*+}_{S0}\\-D^{*\mu+}_{S1}\,\partial_\mu
D^{*-}_{S0})/2+\,ig_1\phi_N(D^{*\mu-}\,\partial_\mu
D^{*+}_0-D^{*\mu+}\,
\partial_\mu D^{*-}_0)/2 +\,ig_1\phi_N(D^{*\mu 0}\,\partial_\mu \overline{D}\,^{*0}_0\\-\overline{D}\,^{*\mu 0}\,\partial_\mu D^{*0}_0)/2
-g_1\phi_N(D^{0\mu}_1\, \partial_\mu
\overline{D}^0+\overline{D}\,^{\mu 0}_1\, \partial_\mu D^0+D^{\mu
+}_1\, \partial_\mu D^-+D^{\mu -}_1\, \partial_\mu D^{+})/2,
\end{multline}
respectively. Note that our Lagrangian is real despite the
imaginary $K_S-K^*$, $D^*_{S0}-D^*_{S1}$, and $D^*_0-D^*$ coupling
because the $K_S-K^*$, $D^*_{S0}-D^*_{S1}$, and $D^*_0-D^*$ mixing
terms are equal to their hermitian conjugates.

In order to obtain canonically normalized fields, we introduce
wave-function renormalization constants labelled $Z_{\eta_{N, S}}$
for $\eta_{N, S}$, $Z_\pi$ for $\overrightarrow{\pi}$, $Z_{K_S}$
for $K_S$, and $Z_K$ for $K$, $Z_{\eta_{C}}$ for $\eta_{C}$,
$Z_{D_S}$ for $D_S$, $Z_{D_{S0}}$ for $D_{S0}$, $Z_{D^*_0}$ for
$D^*_0$, $Z_{D^{*0}_0}$, and $Z_D$ for $D$. We obtain the
following formulas:
\begin{equation}\label{2.13}
Z_{\pi} \equiv
Z_{\eta_N}=\frac{m_{a_1}}{\sqrt{m^2_{a_1}-g_1^2\,\phi^2_N}},\,\,\,\,\,\,\,\,\,Z_{\eta_S}=\frac{m_{f_{1S}}}{\sqrt{m^2_{f_{1S}}-2\,g_1^2\,\phi^2_S}},\,\,\,\,\,\,\,\,\,\,
\end{equation}
\begin{equation}\label{2.14}
Z_{K}=\frac{2m_{K_1}}{\sqrt{4m^2_{K_1}-g_1^2(\phi_N+\sqrt{2}\phi_S)^2}},\,\,\,\,\,\,\,\,\,\,Z_{K_S}=\frac{2m_{K_*}}{\sqrt{4m^2_{K_*}-g_1^2(\phi_N-\sqrt{2}\phi_S)^2}},
\end{equation}
$\\$
as in Ref. \cite{three flavor}, and additionally
\begin{equation}\label{2.15}
Z_{\eta_C}=\frac{m_{\chi_{C1}}}{\sqrt{m^2_{\chi_{C1}}-2g_1^2\phi_C^2}},\,\,\,\,\,\,\,\,\,\,Z_{D_S}=\frac{\sqrt{2}m_{D_{S1}}}{\sqrt{2m^2_{D_{S1}}-g_1^2(\phi_S+\phi_C)^2}},
\end{equation}
\begin{equation}\label{2.16}
Z_{D_{S0}}=\frac{\sqrt{2}m_{D^*_{S}}}{\sqrt{2m^2_{D^*_{S}}-g_1^2(\phi_S-\phi_C)^2}},\,\,\,\,\,Z_{D^*_{0}}=\frac{2\,m_{D^*}}{\sqrt{4m^2_{D^*}-g_1^2(\phi_N-\sqrt{2}\phi_C)^2}},
\end{equation}
\begin{equation}\label{2.17}
Z_{D^{*0}_{0}}=\frac{2\,m_{D^{*0}}}{\sqrt{4m^2_{D^{*0}}-g_1^2(\phi_N-\sqrt{2}\phi_C)^2}},\,\,\,\,Z_{D}=\frac{2\,m_{D_1}}{\sqrt{4m^2_{D_1}-g_1^2(\phi_N+\sqrt{2}\phi_C)^2}},
\end{equation}
$\\$
where $\phi_N=Z_\pi f_\pi$ \cite{denis}, $\phi_{S}=\frac{2
Z_K f_K-\phi_N}{\sqrt{2}}$ for the non-strange and strange
condensates, with $f_\pi=92.4$ MeV and  $f_K=155/\sqrt{2}$ MeV
being the pion and kaon decay constants, respectively, as shown in
Ref.\ \cite{dick}. For the charm condensates, we have
$\phi_C=\frac{1}{\sqrt{2}}(2 Z_D f_D-\phi_N)$ or $\phi_C= \sqrt{2}
Z_{D_S} f_{D_S}-\phi_S$, where $f_D$, and $f_{D_S}$ are the decay
constants of the pseudoscalar $D$ and $D_S$ mesons, respectively.

\section{Parameters and results}

Our model (2.5) has 16 free parameters in the case of $N_f=4$, where
twelve of them have been determined in Ref.\ \cite{dick} for the case $N_f=3$.
Their values are:
\begin{center}%
\begin{tabular}
[c]{|c|c|c|c|}
 \hline Parameter  & Value & Parameter  & Value\\
 \hline  $m_1^2$     & $0.4135\times10^6$ MeV$^2$   & $m_0^2$    & -0.9183$\times10^6$ MeV$^2$\\
 \hline $\delta_N$ & 0     & $\delta_S$ & $0.1511\times10^6$ MeV$^2$\\
 \hline  $g_1$     & 5.8433  & $h_1$      & 0\\
 \hline  $h_2$     & 9.8796 & $h_3$     & 3.8667\\
 \hline  $\phi_N$  & 164.6 MeV& $\phi_S$ & $126.2$ MeV\\
 \hline  $\lambda_1$ & 0   & $\lambda_2$ & 68.2972\\
 \hline
\end{tabular}\\
$\bf Table\, 1:\, Values\,\, of\,\, parameters.$
\end{center}
The parameter $c$ in the axial anomaly term is related to the
corresponding parameter \cite{dick, nf3Giacosa} in the case
$N_f=3$ as follows:
\begin{equation}\label{65}
c=\frac{2\,c_{N_f=3}}{\phi^2_c}\;,
\end{equation}
where $c_{N_f=3}=450.5420\times10^{-6}$ MeV$^{-2}$ from the fit in Ref.\ \cite{dick}.

The three new parameters $\delta_C,\, \phi_C$, and $\varepsilon_C$
related to the mass of the charm quark have been determined by a
fit including the experimental charmed meson masses from the PDG
\cite{PDG} and our equations for the charmed meson masses. Then we
get $\phi_C=198.103\; \rm{MeV},\,\,\, \delta_C=3.742 \times10^6\;
\rm{MeV}^2$, and $\varepsilon_C=1.432\times10^6 \; \rm{MeV}^2$. We
obtain $c=2.296\times 10^{-8} $ in the axial anomaly term
$c\,(\rm{det}\Phi-\rm{det}\Phi^\dagger)^2$. The decay constant of
the isodoublet $D$ is $f_D= 266.52$ MeV, and of the isosinglet
$D_S$ we get $f_{D_S}= 273.8$ MeV, while the experimental values
from the PDG \cite{PDG} are $f_D=206.7$ MeV, and $f_{D_S}= 260.5$
MeV, respectively.

The difference of the square charmed vector and axial-vector
masses are:
\begin{equation}\label{3.2}
m_{D_{S1}}^2-m_{D^*_{S}}^2=2\,(g_1^2-h_3)\phi_S\,\phi_C\;,
\end{equation}
\begin{equation}
m_{D_{1}}^2-m_{D^*}^2=\sqrt{2}\,(g_1^2-h_3)\phi_N\,\phi_C\;,
\end{equation}
and
\begin{equation}
m_{\chi_{C1}}^2-m_{J/\psi}^2=2\,(g_1^2-h_3)\phi^2_C\;.
\end{equation}
These mass differences are interesting because they do not depend on the charm mass. The results for the light mesons are reported in table 2.
 By construction, one finds the same values as in Refs.\ \cite{three flavor, dick}.
\begin{center}%
\begin{tabular}
[c]{|c|c|c|c|} \hline observable &  our value [MeV] & experimental
value [MeV]\\
\hline $m_{f_{1N}}$ & 1186  &1281.8 $\pm$ 0.6\\
\hline $m_{a_{1}}$  & 1185  &1230 $\pm$ 40\\
\hline $m_{f_{1S}}$ & 1372  &1426.4 $\pm$ 0.9\\
\hline $m_{K^*}$    & 885    &891.66 $\pm$ 0.26\\
\hline $m_{K_1}$    & 1281  &1272 $\pm$ 7\\
\hline $m_{\sigma_1}$ & 1362  &(1200-1500)-i(150-250)\\
\hline $m_{a_0}$      & 1363  &1474 $\pm$ 19\\
\hline $m_{\sigma_2}$ & 1531   &1720 $\pm$ 60\\
\hline $m_{\omega_N}$ & 783   &782.65 $\pm$ 0.12\\
\hline $m_{\omega_S}$ & 975   &1019.46 $\pm$\\
\hline $m_\rho$       & 783   &775.5 $\pm$ 38.8\\
\hline $m_{\eta}$     & 509   &547.853 $\pm$ 0.024\\
\hline $m_\pi$        & 141   &139.57018 $\pm$ 0.00035\\
\hline $m_{\eta'}$    & 962   &957.78 $\pm$ 0.06\\
\hline $m_{K_S}$      & 1449   &1425 $\pm$ 50\\
\hline $m_K$          & 485  &493.677 $\pm$ 0.016\\

\hline
\end{tabular}\\
$\bf Table\, 2:$\,\, Light meson masses.
\end{center}
 The results for the (hidden and open) charmed mesons are reported in table 3.
  They have been obtained through a fit to experimental data.
\begin{center}
\begin{tabular}
[c]{|c|c|c|c|} \hline Observable &   Our Value [MeV]  &
Experimental
Value [MeV]\\

\hline $m_{D_{s1}}$    & 2501  &2535.12 $\pm$ 0.13 \\
\hline $m_{D^*_{s}}$   & 2188  &2112.3 $\pm$0.5\\
\hline $m_{D^*}$       & 2155  &2010.28 $\pm$ 0.13\\
\hline $m_{D^{*0}}$    & 2155  &2006.98 $\pm$ 0.15\\
\hline $m_{D_1}$       & 2448  &2421.3 $\pm$ 0.6\\
\hline $m_{\chi_{c1}}$ & 3282  &3510.66 $\pm$ 0.07\\
\hline $m_{\chi_{c0}}$ & 3160  &3414.75 $\pm$ 0.31 \\
\hline $m_{J/\psi}$    & 2911   &3096.916 $\pm$ 0.011\\
\hline $m_{D_0}$       & 1882  &1864.86 $\pm$ 0.13 \\
\hline $m_{\eta_c}$    & 2491  &2981 $\pm$ 1.1\\
\hline $m_{D^*_0}$     & 2416  &2403 $\pm$ 14 $\pm$ 35\\
\hline $m_D$           & 1882  &1869.62 $\pm$ 0.15\\
\hline $m_{D^*_{s0}}$  & 2470  &2317.8 $\pm$ 0.6\\
\hline $m_{D_s}$       & 1900  &1968.49 $\pm$ 0.32\\
\hline $m_{D^{*0}_0}$  & 2416  &2318 $\pm$ 29\\

\hline
\end{tabular}\\
$\bf Table\, 3:$\,\, Open and hidden charmed meson masses.
\end{center}

\section {Conclusions and outlook}
\indent We have extended a linear sigma model with (axial-)vector
degrees of freedom, the so-called
eLSM, to the case of four flavors, $N_f=4$.
The model implements the symmetries of QCD:
the discrete C, P, and T symmetries and the global chiral $U(N_f)_r\times
U(N_f)_l$ symmetry. The latter is broken
spontaneously through the chiral condensate,
explicitly through non-vanishing quark masses, and at the
quantum level through the $U(1)_A$ anomaly. Furthermore, it implements the
dilatation symmetry and its explicit breaking due to the trace anomaly.
In the extension to $N_f=4$ we have
included twelve new charmed mesons in the model, which are the scalar
mesons $ (D^*_0, \,D^*_{s0}, \,\chi_{c0})$, the pseudoscalar
mesons $(D, \,D_s, \,\eta_c)$, the vector mesons
$(D^*, \,D^*_{s}, \,J/\psi)$, and the axial-vector mesons $(D_1,
\,D_{s1}, \,\chi_{c1} )$.  To our
knowledge, this the first time that a model was constructed, which
contains (pseudo)scalars and (axial-)vectors with charm quarks
(open and hidden charmed mesons), and the first time that
the model has been used to describe meson states with high masses.

We have found that we need a new mass term to successfully fit the
masses of charmed mesons.
Implementing spontaneous symmetry breaking
in the model yields not only the known $\eta_N-f_{1N}$,
$\overrightarrow{\pi}$-$\overrightarrow{a}_1$, $\eta_S-f_{1S}$,
$K_S-K^*$, and $K-K_1$ mixings \cite{three flavor} in Eq.\ (2.11)
but also the $\eta_C-\chi_{C1}$, $D_S-D_{S1}$, $D^*_{S0}-D^*_{S1}$,
$D^*_0-D^*$, and $D-D_1$ mixings in Eq.\ (2.12). Removing the
non-diagonal terms in the Lagrangian and subsequently bringing the
kinetic terms of $\eta_{N,S,C},\, \overrightarrow{\pi}, \,K_S, \,K,
\,D_S, \,D_{S0}, \,D^*_0, \,D^{*0}_0$, and $D$ to the canonical form leads us
to define the charmed renormalization coefficients $Z_{\eta_C},\,
Z_{D_S},\, Z_{D_{S0}}, \, Z_{D^*_0},$ $ Z_{D^{*0}_0},$ and $Z_D$.
The squared mass differences $m^2_{D_{S1}}-m^2_{D^*_{S}}, \,
m^2_{D_1}-m^2_{D^*}-m^2$, and $m^2_{\chi_{C1}}-m^2_{J/\psi}$ are,
as seen in Eqs.\ (3.2 - 3.4), independent from
the charm masses, they just depend on the condensates.

The eLSM containing open and hidden charmed mesons has 16 free
parameters. We have fixed 13 of these parameters from the strange
and non-strange sector. The three new unknown parameters have been
fixed in a fit to the expermental values (details will be presented in Ref.\ \cite{charmed}). Moreover, the decay constants of
$D$ and $D_S$ have been calculated. All meson masses in the
Lagrangian (2.5) have been computed, with the same results for
light mesons as in the strange and non-strange sector
\cite{denisthesis, three flavor, dick}, and with (open and hidden)
charmed meson masses being in good agreement with
experimental data \cite{PDG}.

A study in progress is the calculation of the decay widths
of the charmed mesons presented in this work \cite{charmed}.
In the future, we are planning to study the vacuum
phenomenology of the light and heavy tetraquarks and the inclusion of the scalar and pseudoscalar glueballs in $N_f=4$.

\section*{Acknowledgments}

The author thanks F.\ Giacosa, D.\ Parganlija, and D.H.\ Rischke for
cooperation and  useful discussions. She acknowledges
support from DAAD and HGS-HIRe.

\end{document}